\documentclass[traditabstract]{aa}
\usepackage{txfonts}
\usepackage{graphicx}
\usepackage{color}
\usepackage{CJK} 
\usepackage{natbib}

\newcommand\phs{\phantom{$-$}}
\newcommand{\rb}[1]{\raisebox{1.5ex}[-1.5ex]{#1}}

\begin{document}

\begin{CJK}{UTF8}{gbsn}

\title{
Chemical composition of the outer halo globular cluster Palomar\,15
}

\author{Andreas Koch\inst{1} 
  \and Siyi Xu (许\CJKfamily{bsmi}偲\CJKfamily{gbsn}艺)\inst{2}
  \and R. Michael Rich\inst{3}
  }
  
\authorrunning{A. Koch et al.}
\titlerunning{Chemical abundances in Pal\,15}
\offprints{A. Koch;  \email{andreas.koch@uni-heidelberg.de}}

\institute{
Zentrum f\"ur Astronomie der Universit\"at Heidelberg, Astronomisches Rechen-Institut, M\"onchhofstr. 12, 69120 Heidelberg, Germany 
  \and Gemini Observatory, Northern Operations Center, 670 N. A'ohoku Place, Hilo, HI 96720, USA
  \and University of California Los Angeles, Department of Physics \& Astronomy, Los Angeles, CA, USA
    }
\date{}
\abstract {
Globular clusters (GCs) in the outer Milky Way halo are important tracers of the assembly history of our Galaxy. 
Only a few of these objects show spreads in heavier elements beyond the canonical light-element variations that have essentially been found
throughout the entire Galactic GC system, suggesting a more complex origin and 
evolution of these objects. 
Here, we present the first abundance analysis of three red giants in the remote ($R_{\rm GC}=38$ kpc) outer halo GC Palomar 15, based on medium-resolution spectra obtained with the Keck/ESI instrument.
Our results ascertain a low iron abundance of $-$1.94$\pm$0.06 dex with 
no evidence of any significant abundance spreads, although this is based on low number statistics.
Overall, abundance ratios of 16 species were measured, including carbon, {Na, Al}, $\alpha$-peak (Mg,Si,Ca,Ti) and Fe-peak (Sc,V,Cr,Fe,Co,Ni) elements, and the 
three neutron-capture elements Sr, Ba, and Eu. 
The majority of abundances are compatible with those of halo field stars and those found in other GCs in the outer and inner halos at similar metallicity.
{Pal~15 is enhanced to [Mg/Fe]=0.45 dex, while other $\alpha$-elements, Ca and Ti, are lower by 0.3 dex.
Taking Mg as a representative for [$\alpha$/Fe], and coupled with the lack of any significant spread in any of the studied 
elements we conclude that Pal~15 is  typical } of the 
outer halo, as is bolstered by its chemical  similarity to the benchmark outer halo cluster NGC~7492.
{One star shows evidence of elevated Na and Al abundances, hinting at the presence of multiple stellar populations in this cluster.}}
\maketitle 
%
%
%
%
%
%
\section{Introduction}
Globular clusters (GCs), in particular those at large Galactocentric radii, are palmary objects for tracing the 
formation and structure of the Milky Way (MW).
An inner--outer dichotomy as seen in various stellar tracers in the MW halo \citep{SearleZinn1978,Hartwick1987,Carollo2007} is also present in the nearby galaxy 
M31 \citep{Koch2008M31} and predicted by models of structure formation \citep{Cooper2013,Pillepich2015}.
Here the outer halo GCs are vital study cases since 
some of these remote systems tend to be a few Gyr younger    than inner halo clusters of the same metallicity, as is  reflected, for instance, in the morphologies 
of their horizontal branches \citep{Stetson1999,Catelan2001,Marin-Franch2009}. 
This diversity in ages has often been interpreted as a sign of their accretion origin. 
Other distant GCs are, however, coeval with their old, metal-poor inner halo counterparts \citep{Dotter2011}, and many of the 
outer halo clusters are also chemically indistinguishable from 
the inner ones \citep{Koch2009,Koch2010}.

The relevance of the effects within GCs has gained  
major attention through their complex stellar populations (e.g., \citealt{Gratton2012}, and references therein) 
and the presence of pronounced light-element variations  
\citep[e.g.,][]{Kayser2008,Carretta2009NaO}, all of which have caused a change in  the canonical picture of ``simple'' GC formation
 (as comprehensively reviewed by \citealt{Bastian2018}).
Spreads in iron and some heavy chemical elements, however, have so far only been reported for a minority of mainly massive  stellar systems, which 
suggests a different origin, possibly as parts of former dwarf galaxies \citep[e.g.,][]{Mucciarelli2012,Kacharov2013,Marino2015,Johnson2017,PiattiKoch2018}. 

Here we report on the first chemical abundance study of red giant branch stars in the outer halo GC 
Palomar 15 (herafter Pal~15).
Pal~15 
 is an old GC (13$\pm$0.5 Gyr; \citealt{Dotter2011}) situated at a Galactocentric radius of 38.4 kpc, 
making it one of the most remote GC satellites to the MW. 
Only ten GCs in the catalog of Galactic GCs of \citet[][2010 version]{Harris1996} are located at even further distances. 
While previous studies at low spectral resolution asserted a low metallicity of $-$2 dex \citep{DaCosta1995}, no detailed 
chemical abundances have been determined in this object. 

This paper is organized as follows. In Sect.~2 we introduce the data acquisition and reduction. We describe the ensuing abundance analysis   in Sect.~3, before turning to the results in Sect.~4. In Sect.~5 we compare our findings for Pal~15 with other outer halos GCs beyond 20 kpc, and we 
summarize our findings in Sect.~6. 
\section{Observations and data reduction}
As Pal~15 is a remote, faint system we selected the brightest red giant candidates from the Hubble Space Telescope (HST) photometry of \citet{Dotter2011}, complemented by infrared magnitudes from the Two Micron All Sky Survey \citep[2MASS;][]{Cutri2003}. 
The respective color-magnitude diagram (CMD) is shown in Fig.~1, and in Table~1 we summarize the targets' main properties.
Here we note that 
all our targets are too faint to be extracted from Gaia DR2 \citep{GaiaDR2} so that no useful parallaxes could be extracted. 
\begin{figure}[htb]
\centering
\includegraphics[width=0.51\hsize]{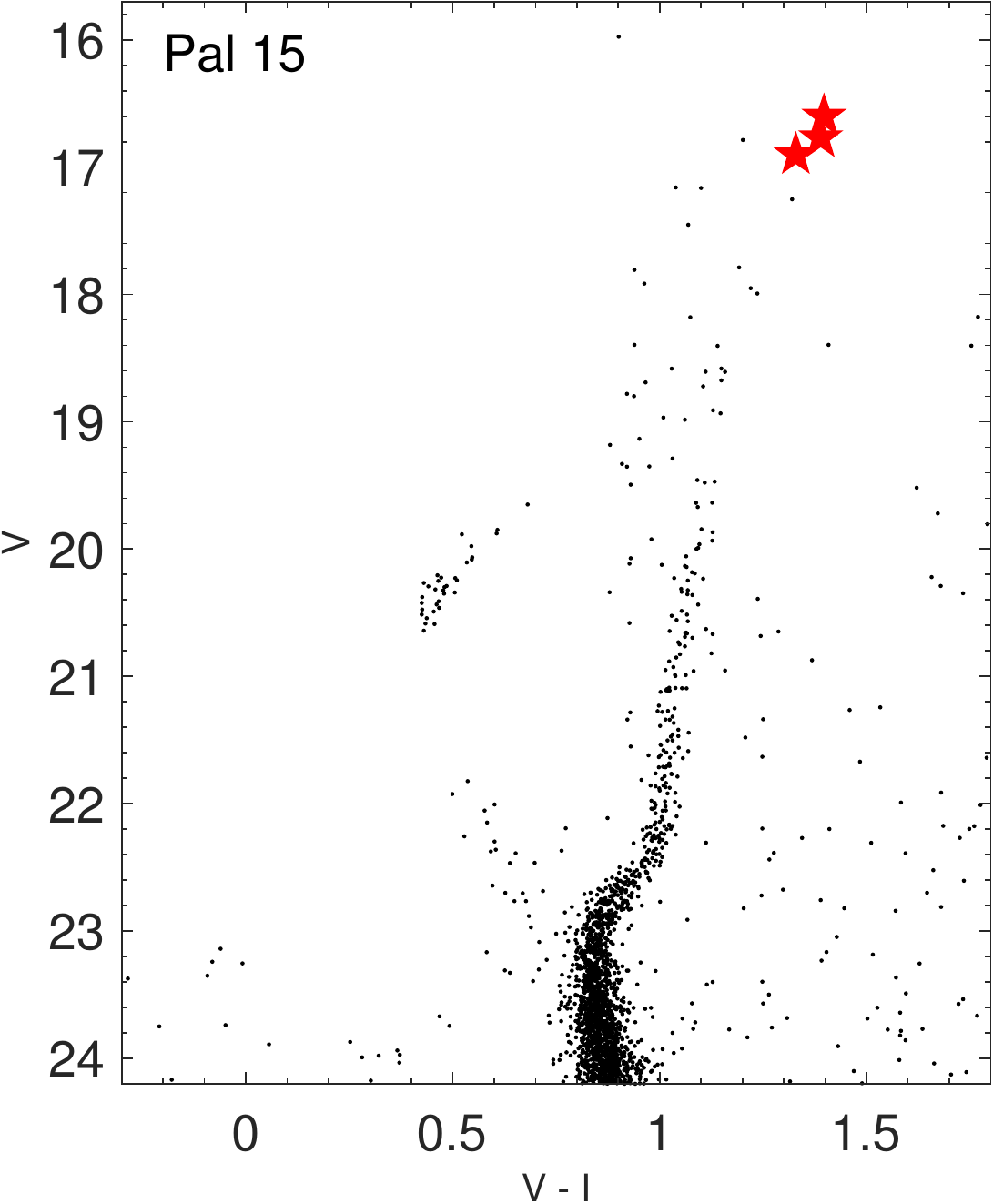}
\includegraphics[width=0.48\hsize]{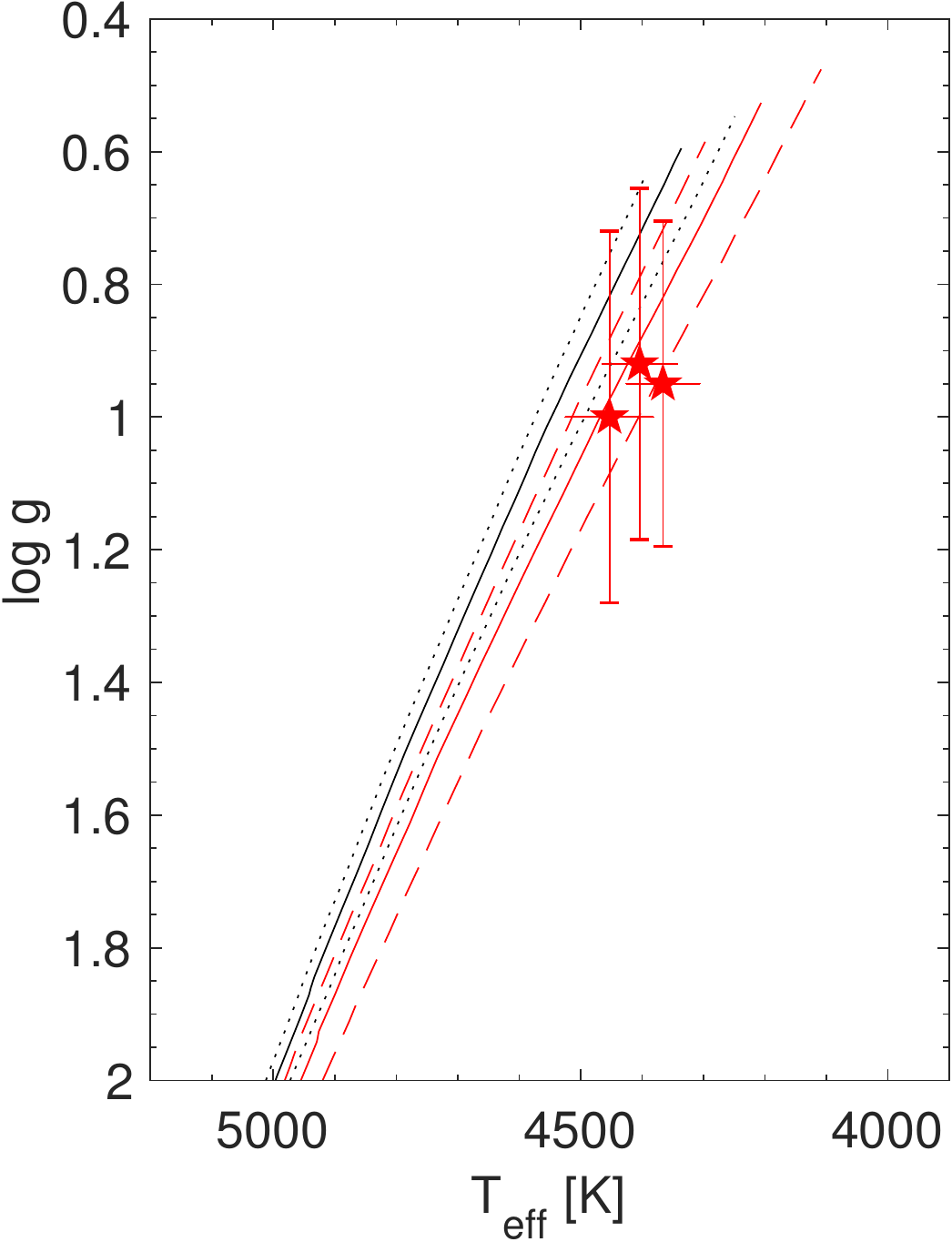}
\caption{{\em Left panel}: CMD  of the target stars using the HST photometry of \citet{Dotter2011}. 
{\em Right panel}: Kiel diagram with the spectroscopic parameters and old (12.4 Gyr), metal-poor ($-2.2$, $-2.0$, $-1.8$), 
Dartmouth isochrones \citep{Dotter2008} {with $\alpha$-enhancement (red lines) and without (black lines)}.
Our targets are indicated as red stars.}
\end{figure}
\begin{table*}[htb]
\caption{Details of targets and observations.}             
\centering          
\begin{tabular}{rccccccccc}     
\hline\hline       
& Alternate & $\alpha$  & $\delta$ & V & I & K  & t$_{\rm exp}$ & S/N & v$_{\rm HC}$  \\
 \rb{ID\tablefootmark{a}} & ID\tablefootmark{b} & (J2000.0) & (J2000.0)  & [mag] & [mag] & [mag] &   [s] & [pixel$^{-1}$]  & [km\,s$^{-1}$] \\
\hline
\phantom{1}3584    & SC-3 & 16:59:52.15   & $-$00:33:23.14  &  16.601 & 15.204   & 12.969   & 3$\times$900 & 55  &   72.1$\pm$0.6       \\
\phantom{1}8469    & SC-4 & 16:59:51.90   & $-$00:32:53.81  &  16.770 & 15.381   & 13.281   & 3$\times$1200 & 73 &   74.7$\pm$0.7      \\          
11726  & SC-7 & 16:59:56.05   & $-$00:32:23.23  &  16.902 & 15.573   & 13.462   & 3$\times$1300 & 73   &    70.3$\pm$0.8 \\      
\hline
\hline                  
\end{tabular}
\tablefoot{
\tablefoottext{a}{Identifications from \citet{Dotter2011}.}
\tablefoottext{b}{Cross-identifications from  \citet{Peterson1989}.}}
\end{table*}

Our observations were carried out during the night of May 17, 2013, using the Echellette Spectrograph and Imager (ESI)  instrument \citep{Sheinis2002} at the Keck\,II telescope at Mauna Kea, Hawai'i. 
We employed  a slit width of 0.5$\arcsec$, yielding an intermediate resolving power of $R$=8000. 
Our data were reduced in a similar manner to \citet{Hansen2016RR}, by 
using the ESI-specific package within the MAKEE\footnote{MAKEE was developed by T. A. Barlow specifically for reduction of Keck HIRES and ESI data. 
It is freely available on the World Wide Web at the Keck Observatory home page, \tt{http://www2.keck.hawaii.edu/inst/hires/makeewww}} 
data reduction tool. 
As a result, the final spectra cover a full wavelength range of 4550--9000\AA~and their signal-to-noise ratios (S/N)  are in the range of 55--75 per pixel across the full spectrum. 

All targets have already been classified as members in the kinematic study of \citet{Peterson1989}, which we confirmed here 
by our radial velocity measurements using {\sc iraf}'s {\em fxcor} tool against a synthetic red giant star template. 
For the three Pal~15 stars we find a mean heliocentric velocity and velocity dispersion of 
72.4$\pm$1.0 km\,s$^{-1}$ and
1.6$\pm$0.8 km\,s$^{-1}$. 
These values are consistent with the findings of  \citet{Peterson1989}, and the latter, lower value is fully in line with the GC's low mass at $M_V=-5.5$ \citep{Peterson1989,Pryor1993}. 
\section{Abundance analysis}
Given the low resolution of our spectra, detailed equivalent width measurements and the ensuing 
excitation and ionization balances are inadequate for an accurate 
stellar parameter determination. 
Thus, we employed a two-step process for the abundance measurements. 
\subsection{SP\_Ace (T$_{\rm eff}$, log\,$g$, [Fe/H], [Mg,Si,Ca,Sc,Ti,V,Cr,Co,Ni/Fe])}
First we used the SP\_Ace code \citep{Boeche2016} 
to obtain the stars' effective temperatures and surface gravities using empirical curves of growth for a vast number of 
atomic lines across the entire spectral range.
From these parameters the microturbulence was extracted by using the empirical formula provided in \citet{Boeche2016}. 
Furthermore, SP\_Ace returns Fe abundances and chemical element ratios of several $\alpha$-elements (Mg,Si,Ca,Ti) and Fe-peak species (Sc,V,Cr,Co,Ni).
The typical temperature errors as returned by SP\_Ace are 80 K and the surface gravity could be determined to within 0.25 dex, 
leading to a typical  uncertainty on the microturbulence of 0.25 kms\,s$^{-1}$. 
{Color-temperature calibrations such as  the (V-K) relations of \citet{RamirezMelendez2005}
return T$_{\rm eff}$ values that are $\sim$130 K warmer  than our values,  with a 1$\sigma$ deviation of 120 K. This fair agreement reinforces 
our use of  SP\_Ace for deriving the stellar parameters of our sample.}

All best-fit parameters are listed in Table~2, while Table~3 contains the abundance ratios determined by SP\_Ace. 
Here  we list the abundances as [X/Fe] together with the 1$\sigma$ line-to-line scatter returned by SP\_Ace, and the numbers of lines used in the analysis.
\begin{table}[htb]
\caption{Stellar parameters of the individual target stars based on SP\_Ace.}             
\centering          
\begin{tabular}{cccc}     
\hline\hline       
Parameter  &  {11726}& {8469} & {3584} \\
\hline
T$_{\rm eff}$ [K] & 4453$\pm$140 & 4404$\pm$118 & 4366$\pm$115   \\
log\,$g$ & \phs1.00$_{-0.37}^{+0.19}$ & \phs0.92$_{-0.39}^{+0.14}$ & \phs0.95$_{-0.31}^{+0.18}$   \\
$\xi$ [km\,s$^{-1}$] & 1.95$\pm$0.13 & 1.92$\pm$0.11 & 1.88$\pm$0.11  \\
\hline
\hline                  
\end{tabular}
\end{table}
\begin{table*}[htb]
\caption{Abundance ratios of the individual target stars and the GC mean.}             
\centering          
\begin{tabular}{crcrcrcrcrcrcrcr}     
\hline\hline       
Abundance & \multicolumn{3}{c}{11726}&&  \multicolumn{3}{c}{8469} &&  \multicolumn{3}{c}{3584} && \multicolumn{2}{c}{Pal 15}  &  \\
\cline{2-4}\cline{6-8}\cline{10-12}\cline{14-15}
{ratio\tablefootmark{a}} & [X/Fe] & $\sigma$ & N  && [X/Fe] & $\sigma$ & N  && [X/Fe] & $\sigma$ & N &&  $<$[X/Fe]$>$ & $\sigma$ & \rb{Note}\\
\hline
$[$Fe/H]  & $-$1.94 & 0.05 & 355 && $-$1.99 & 0.05 & 352 && $-$1.90 & 0.04 & 344 &&  $-$1.94$\pm$0.02 & 0.04$\pm$0.02 & SP\_Ace   \\
$[$C/Fe]  & $-1.11$ & $^{+0.22}_{-0.34}$  & \phantom{11}1  && $-1.41$ & $^{+0.12}_{-0.15}$  & \phantom{11}1    && $-0.89$ & $^{+0.27}_{-0.40}$ &  \phantom{11}1      &&  $-$1.19$\pm$0.25 &  $<$0.01      & Synth \\
$[$Na/Fe]\tablefootmark{b} & \phs0.66 & 0.01 & \phantom{11}2 &&  \phs0.26 & 0.13  & \phantom{11}2 &&  \phs0.10 & 0.16 & \phantom{11}2  && \phs0.36$\pm$0.14 &        0.23$\pm$0.10      & Synth  \\
$[$Mg/Fe] & \phs0.41 & 0.21 & \phantom{11}5 &&  \phs0.47 & 0.18  & \phantom{11}6 &&  \phs0.48 & 0.19 & \phantom{11}4  && \phs0.45$\pm$0.18 &      $<$0.01      & SP\_Ace  \\
$[$Al/Fe] &  \phs0.70 & 0.30  & \phantom{11}4 &&  \phs0.30 & 0.30  & \phantom{11}4 &&  $<$0.50 & \dots & \phantom{11}4  && \phs0.50$\pm$0.17 &      $<$0.01      & Synth  \\
$[$Si/Fe]\tablefootmark{c} & $-$0.04\rlap{:} & 0.18  & \phantom{1}17 && $-$0.16\rlap{:} & 0.23 & \phantom{1}15    && $-$0.15\rlap{:} & 0.20 & \phantom{1}15     &&  $-$0.11$:\pm$0.03 & 0.03$\pm$0.04    & SP\_Ace  \\
$[$Ca/Fe] & \phs0.13 & 0.07 & \phantom{1}26 &&  \phs0.12 & 0.06 & \phantom{1}26  &&  \phs0.17 & 0.05 & \phantom{1}26   && \phs0.14$\pm$0.01 & 0.02$\pm$0.01    & SP\_Ace  \\
$[$Sc/Fe] & \phs0.00 & 0.11 & \phantom{1}15 &&  \phs0.04 & 0.13 & \phantom{1}15  &&  \phs0.09 & 0.09 & \phantom{1}14   && \phs0.05$\pm$0.02 & 0.03$\pm$0.02    & SP\_Ace  \\
$[$Ti/Fe] & $-$0.01 & 0.10 & \phantom{1}58  &&  \phs0.01 & 0.10 & \phantom{1}54  &&  \phs0.11 & 0.07 & \phantom{1}56   && \phs0.04$\pm$0.03 & 0.05$\pm$0.02    & SP\_Ace  \\
$[$V/Fe]  & $-$0.23 & 0.15 & \phantom{1}22  && $-$0.26 & 0.23 & \phantom{1}24    && $-$0.25 & 0.17 & \phantom{1}28     &&  $-$0.23$\pm$0.03 &  $<$0.01       & SP\_Ace  \\
$[$Cr/Fe] & $-$0.08 & 0.19 & \phantom{1}24  && $-$0.07 & 0.18  & \phantom{1}24    && $-$0.11 & 0.15 & \phantom{1}23     &&  $-$0.09$\pm$0.17 &  $<$0.01       & SP\_Ace  \\
$[$Co/Fe] & $-$0.14 & 0.20 & \phantom{1}20  && \phs0.01 & 0.28 & \phantom{1}24    && $-$0.04 & 0.18 & \phantom{1}23     &&  $-$0.04$\pm$0.04 &  $<$0.01       & SP\_Ace  \\
$[$Ni/Fe] & $-$0.11 & 0.09 & \phantom{1}62  && $-$0.08 & 0.08 & \phantom{1}60    && $-$0.04 & 0.06 & \phantom{1}58     &&  $-$0.08$\pm$0.02 & 0.03$\pm$0.01    & SP\_Ace  \\
$[$Sr/Fe] & \phs0.60 & 0.15 & \phantom{11}1 && \phs0.55 & 0.20 & \phantom{11}1 && \phs0.40 & 0.25 & \phantom{11}1 && \phs0.55$\pm$0.20 & $<$0.01 & Synth  \\                        
$[$Ba/Fe] & \phs0.54 & 0.19 & \phantom{11}4 && \phs0.45 & 0.25 & \phantom{11}4 && \phs0.62 & 0.27 & \phantom{11}4 && \phs0.54$\pm$0.10 & $<$0.01 & Synth \\
$[$Eu/Fe] & \phs0.40 & 0.20 & \phantom{11}2 && \phs0.40 & 0.40 & \phantom{11}2  && \phs0.50 & 0.40 & \phantom{11}2  && \phs0.42$\pm$0.13 & $<$0.01  & Synth \\ 
\hline
\hline                  
\end{tabular}
\tablefoot{
\tablefoottext{a}{All abundance ratios are stated as the mean value, 1$\sigma$ standard deviation, and number of lines used. [C/Fe] was measured from the molecular
G band. Abundance errors from spectral synthesis are based on a weighted average after visual quality control of the individual lines.}
\tablefoottext{b}{NLTE value.}
\tablefoottext{c}{Uncertain values.}
}
\end{table*}
Several examples of spectral regions around absorption lines of $\alpha$-elements 
used by SP\_Ace are shown in Fig.~2 with the best synthesis  in comparison with the observed spectrum of star 11726 . 
\begin{figure}[htb]
\centering
\includegraphics[width=1\hsize]{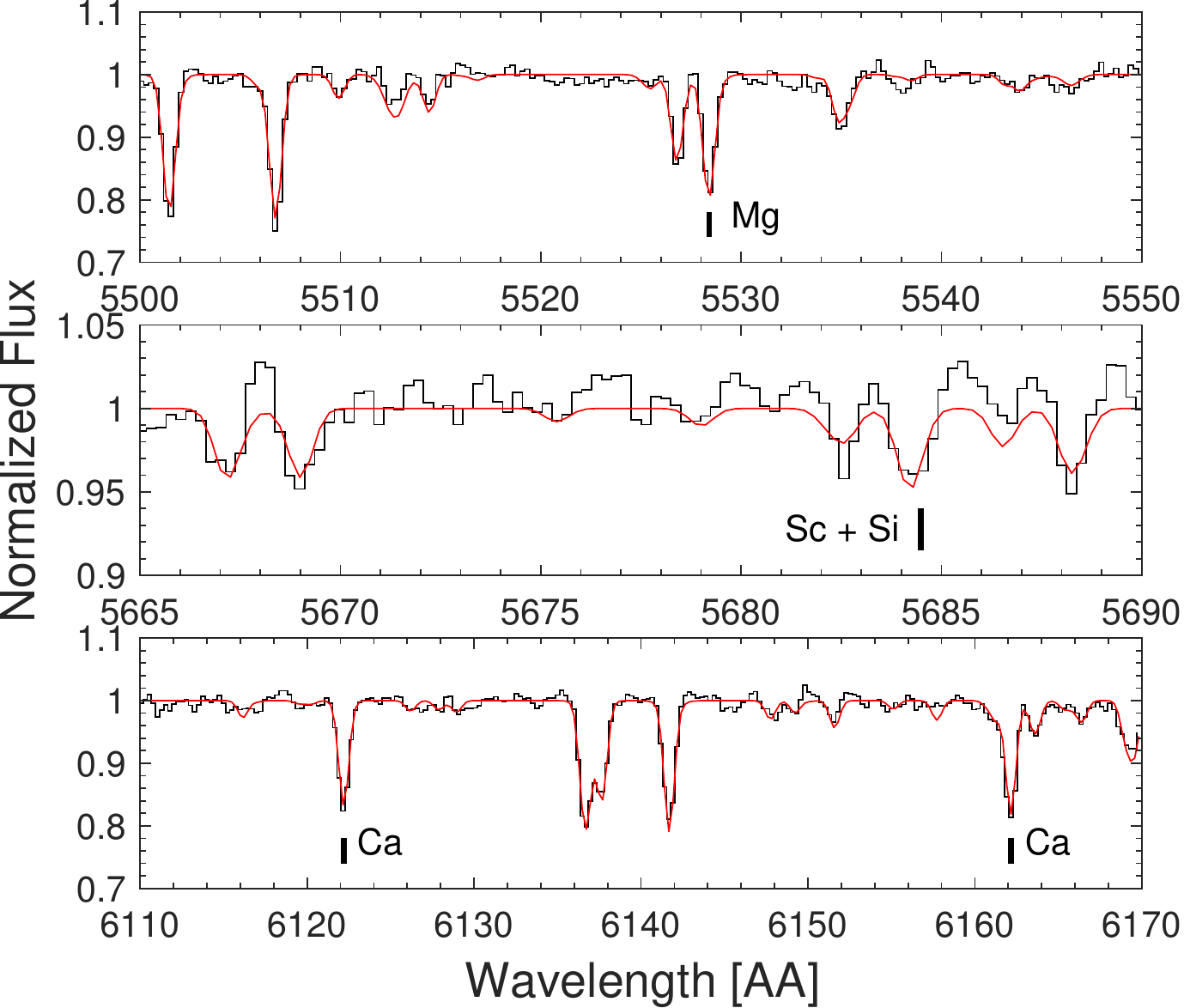} 
\caption{Sample regions around $\alpha$-element lines that were used by SP\_Ace. The best model computed by SP\_Ace is 
shown in red superimposed on  the observed spectrum of star 11726.t}
\end{figure}

\subsection{Spectral synthesis with MOOG ([C,Sr,Ba,Eu/Fe])}
The stellar parameters derived above were then used to create model atmospheres by interpolating Kurucz's  1D  72-layer, plane-parallel, line-blanketed model grid without convective overshoot\footnote{http://kurucz.harvard.edu/grids.html}, where we assumed 
that local thermodynamic equilibrium holds for all species. 
This model grid further incorporated the $\alpha$-enhanced opacity distributions, AODFNEW \citep{CastelliKurucz2004}, as prompted  by 
the $\alpha$-enhancement of the stars measured by SP\_Ace.

For all further analyses, we used the 2014 version of the stellar abundance code MOOG \citep{Sneden1973}
to determine abundances of a number of species via spectral synthesis. 
The line list for this purpose is based on \citet{Koch2016} with further additions from 
\citet{CJHansen2012} and \citet{CJHansen2013}.
All abundances thus determined were placed on the solar, photospheric scale of \citet{Asplund2009}.
This procedure  was  proved successful on medium-resolution ESI-spectra by \citet{Lai2009}.
\section{Results}
Table~3 lists the abundance ratios we measured either from SP\_Ace  or from our spectral syntheses, where the last column
details the method used for abundance determination. 
\subsection{Iron}
Previous studies have established the metal-poor nature of Pal~15 by CMD fitting \citep{Dotter2011} and from low-resolution CaT spectroscopy  
\citep{DaCosta1995}, both of which found an [Fe/H] of $-$2 dex. This is confirmed by our spectroscopic study.
From our three stars, we find a mean Fe-abundance of $-$1.94$\pm$0.06 with a 1$\sigma$ dispersion of 0.09$\pm$0.11 dex  determined in a 
maximum-likelihood approach. This low value is in line with the overall low metallicities of GCs in the outer halo, which peak at a 
mean [Fe/H] of $-1.7$ dex beyond 20 kpc. In turn, about three objects at these large distances lie toward the lower cutoff
of the GC metallicity distribution below $-$2 dex. Thus, Pal~15 is a typical representative of the outer halo population. 

To date, Fe spreads have  chiefly been found in rather massive GCs, which may point to more massive progenitors such as dwarf galaxies, 
while on the order of 5\% of old galactic GC systems  show significant intrinsic Fe spreads in excess of 0.05 dex \citep{Johnson2015,Marino2015,PiattiKoch2018}. 
Considering its fairly low mass, at $M_V = -5.5$ mag, Pal~15's lack of a significant Fe spread argues in favor of it being a regular GC with an ordinary enrichment history. 
In the following, we  investigate this further by using additional tracers of chemical evolution.
\subsection{Carbon}
Carbon abundances were derived from spectral synthesis of the CH G band at 4300\AA; the typical fits (as determined in a $\chi^2$ sense) are shown in Fig.~3. 
\begin{figure}[htb]
\centering
\includegraphics[width=0.8\hsize]{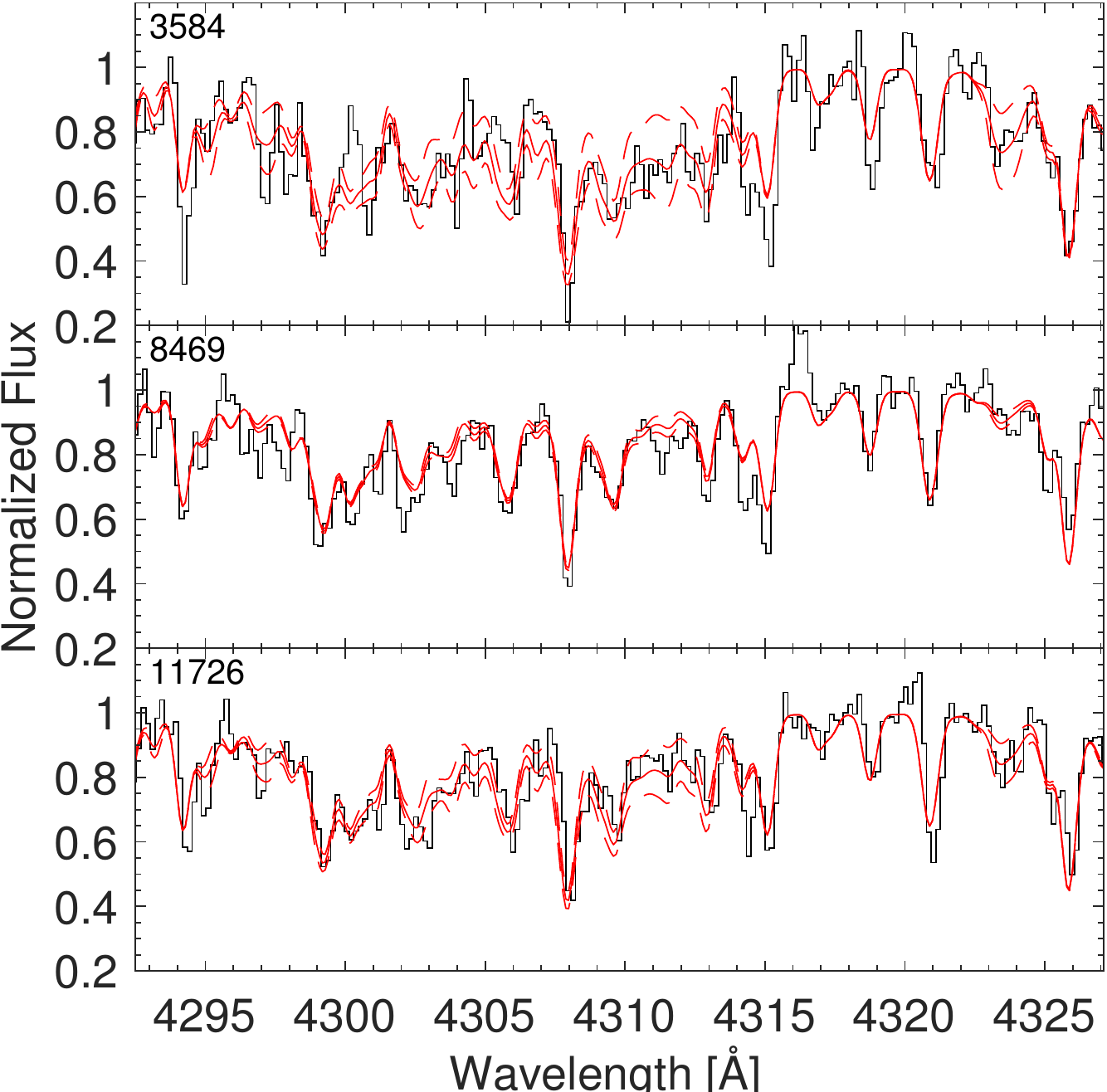}
\caption{Spectra of the three red giants near the CH G band. The red lines show the best-fit spectrum and the synthesis of the error margins as dashed lines.}
\end{figure}

The results are shown in Fig.~4 in comparison with data for red giants in metal-poor ([Fe/H]$<-2$) GCs from \citet{Kirby2015}. 
In all three stars of our study, the carbon abundance are very low, as is expected for such luminous red giants due to  deep mixing that occurs
during stellar evolution \citep{Spite2005,Placco2014,CJHansen2016}.  Correcting for evolutionary effects  would lead to increased values for the stars'
intrinsic carbon abundances, but given the regularity of our results we do not pursue the investigation of carbon  further. 
\begin{figure}[htb]
\centering
\includegraphics[width=1\hsize]{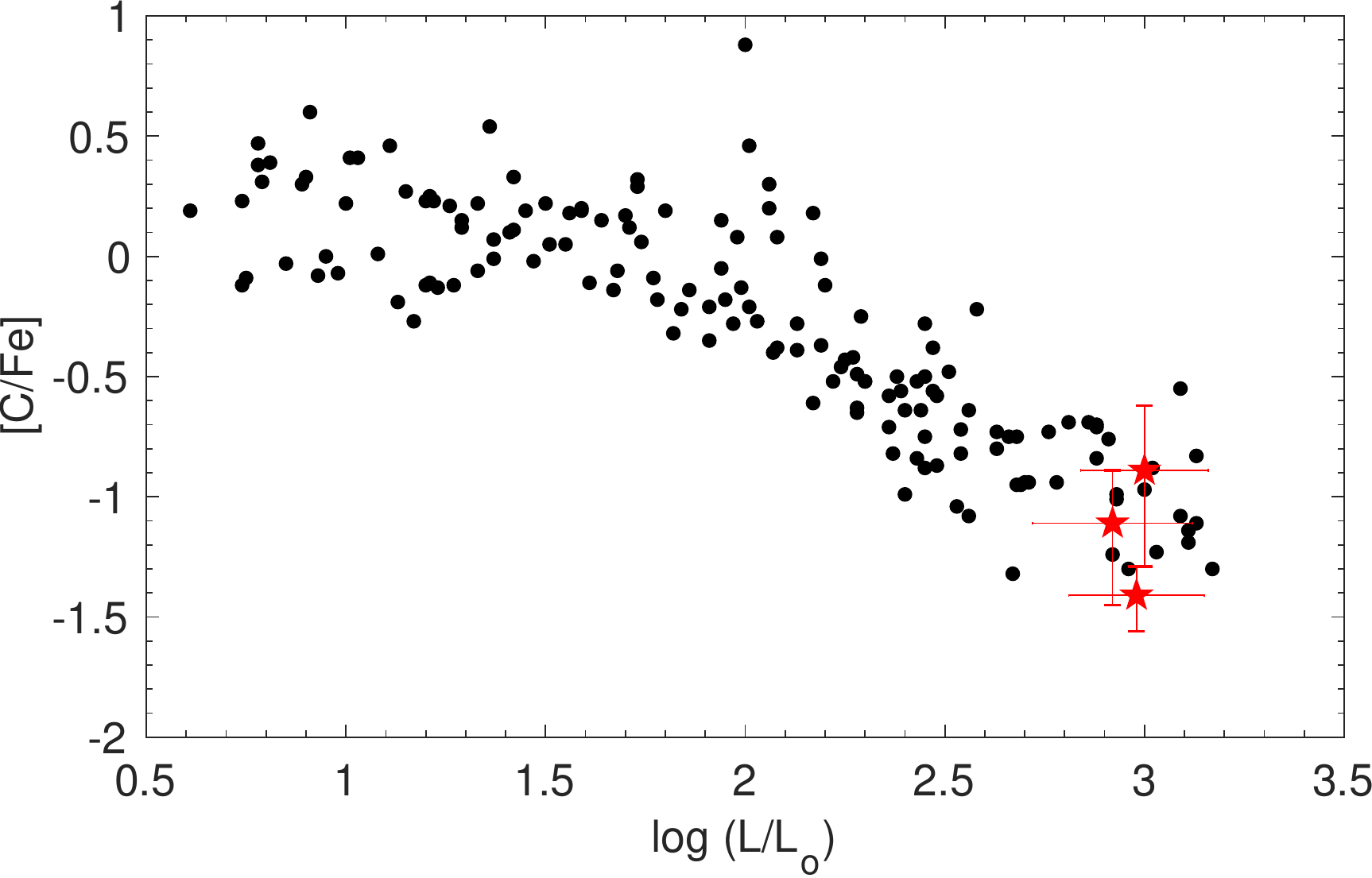} 
\caption{Carbon abundances of red giants in Pal~15 (red stars) and metal-poor halo GCs (black dots; \citealt{Kirby2015}).The latter 
are not corrected for evolutionary effects in this representation.}
\end{figure}
\subsection{Light elements (Na, Al)}
{Sodium abundances were measured from the strong Na D resonance lines that we carefully deblended from interstellar contamination. 
These lines are very strong in our stars, with equivalent widths in excess of 300 m\AA, so we advise caution with regard to the following interpretations. 
The resulting Na abundances were corrected for departures from local thermodynamic equilibrium (LTE) \citep{Lind2011} and Table~3 lists the final non-LTE (NLTE) abundances. 

One of the key signatures of GCs is a pronounced anti-correlation between Na and O, indicating the presence of several stellar generations in each cluster.
This also leads to the presence of light-element spreads that contrast the lack of any such variations in the heavy elements, safe for a few exceptions
at the higher mass tail of GCs. Our spectral quality did not permit us to derive any oxygen abundances;  for stars of our given stellar parameters, the 
commonly accessible O lines at 6300 or 7770\AA~are too weak to be detected. 

Two of the stars in our sample have moderate enhancements in Na that are fully consistent with the properties of the first halo-like 
generation that formed in the GC \citep{Carretta2009NaO}. In turn, the brightest star shows a considerably higher abundance 
of $\sim$0.65 dex, which could indicate that it is a member of the second generation of Na-rich, O-poor stars, 
which also suggests that  Pal 15 hosts multiple stellar generations as are seen in other GCs of similar age and mass (see, e.g., Fig.~9 in 
\citealt{Bragaglia2017}).

We synthesized the Al lines at 6696, 6698\AA, and also obtained an estimate from the doublet at 8872\AA, but since they
are very weak transitions, we urge caution when using the following results. 
We merely point out that star 11726, which shows the highest Na abundance in our sample, also has an elevated [Al/Fe] of $\sim$0.7 dex.
The other two stars (also poorer in Na) have more moderate Al abundances, which provides further evidence that  Pal~15 also shows
light-element variations due to the presence of multiple populations \citep{Carretta2018}. 
}%
\subsection{$\alpha$-elements (Mg, Si, Ca, Ti)}
The abundances of the key $\alpha$-elements are based on our analysis with SP\_Ace
and have been  verified for the brightest star, with MOOG.  
Our results are shown in Fig.~5, where we also overplot metal-poor halo field stars from \citet{Roederer2014} 
and the disk sample of \citet{Bensby2014}. Furthermore, we show the mean abundances and abundance spreads of 
several outer halo GCs from the literature, which we  discuss in more detail in Sect.~5. 
\begin{figure}[htb]
\centering
\includegraphics[width=1\hsize]{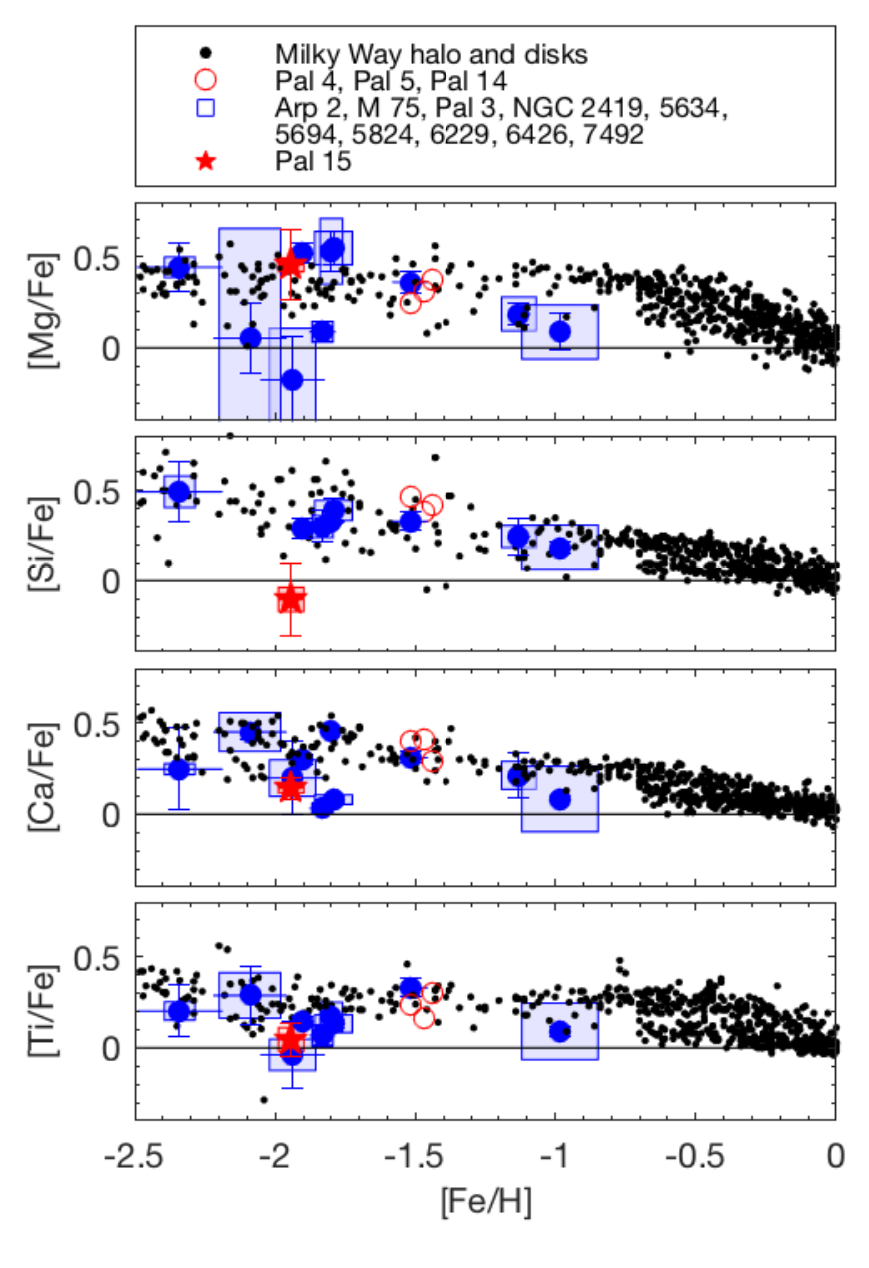}  
\caption{Abundances of the $\alpha$-elements for Pal~15 (red star) and the reference outer halo GCs. 
Here, red open circles show GCs with mean abundances derived from co-added spectra.
The blue boxplots refer to the observed 1$\sigma$ spread in a given GC (extent of the boxes in either dimension), while 
the inlaid error bars depict the mean abundance error on the respective measurements.}
\end{figure}

By grouping together all four elements, we report a straight mean [$\alpha$/Fe] of 0.13$\pm$0.01 dex; this is  a mild enhancement that is almost compatible with solar abundances.  
It is a rather low value for  low-metallicity GCs, which  typically  follow the enrichment of the  underlying Galactic component 
\citep[e.g.,][]{Pritzl2005,Hendricks2016}. Systematically lower values of [$\alpha$/Fe] with respect to halo field stars 
are a trademark signature of the low-mass dwarf galaxies \citep[e.g.,][]{Matteucci1990,Shetrone2001,Koch2009Review,Tolstoy2009}. This has spawned
the notion that GCs that show depleted $\alpha$-abundances could be accreted objects that  formerly belonged to more massive, complex systems  
such as these dwarfs \citep[e.g.,][]{Brown1997,Cohen2004,Villanova2013}

However,  this simplistic combination of individual $\alpha$-elements is intrinsically problematic due to the different  (explosive versus hydrostatic)  burning 
phases producing these elements, leading to distinctive behaviors \citep[e.g.,][]{Woosley1995,Koch2011}. 
We note the  significantly elevated [Mg/Fe] ratio of 0.46$\pm$0.05 that lies on the plateau of halo field stars.
This contrasts a strongly depleted [Si/Fe] of $-$0.10$\pm$0.03.
The reason for this lies  in the weakness of the Si lines (typically below 16 m\AA) employed by SP\_Ace
coupled with the intermediate resolving power of our spectra. 
Here, we note that the code does not contain transitions in the wavelength range of 6860--8400\AA, whereas redder wavelengths up to 8800\AA~are again considered.
Therefore,  this overinterpretation of the low Si abundance should be taken with a grain of salt. 

The explosive element abundances for Ca and Ti take values in between the enhanced (hydrostatic) 
Mg and depleted Si, and they lie toward the lower boundaries outlined by the halo field stars. 
Finally, we note that \citet{Dotter2011} found that an $\alpha$-enhancement of 0.4 dex in their isochrones best represents the cluster's CMD (right panel of Fig.~1).
Taking Mg as a prime $\alpha$-element that, as a significant electron donor,  strongly affects stellar atmospheres and therefore the appearance of the isochrones, 
the concordance between our reported [Mg/Fe] and the CMD fitting corroborates the enhancement of Pal~15 and its place
{on the Mg plateau that is seen for the majority of halo stars and GCs.} 
\subsection{Fe-peak elements (Sc, V, Cr, Co, Ni)}
Given the fortuitous spectral range of our data, we were able to extract abundance ratios for five Fe-peak elements using 
SP\_Ace. They are indicated in Fig.~6 in comparison with the same halo data as mentioned above, with added measurements in the 
Galactic disk from  \citet{Battistini2015}.
\begin{figure}[htb]
\centering
\includegraphics[width=1\hsize]{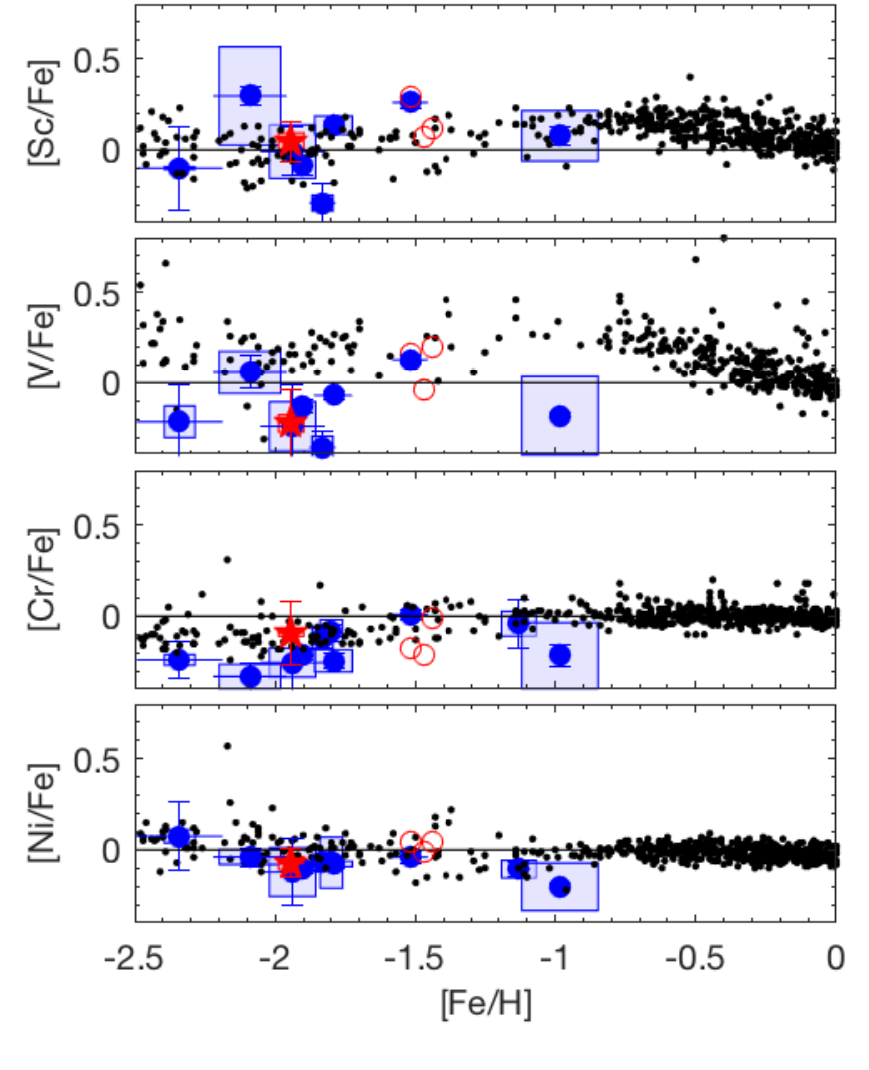} 
\caption{Same as Fig.~5, but for the Fe-peak elements Sc, V, Cr, and Ni.}
\end{figure}

The Fe-peak elements are synthesized in  supernovae (SNe) of both Type Ia  and II, while at higher,  solar metallicity, 
there is a preponderance of SNe Ia in the production of these elements \citep[e.g.,][]{Woosley1995,Kobayashi2006}. 
There was no surprise from the abundance ratios we derived in Pal~15 stars in 
comparison with halo stars of similarly low metallicity, and we note a 
remarkable overlap of our GC stars with the abundances of the stellar halo. 
We note, however,  that none of our results was corrected for departures from LTE, 
which would mainly have some effects on Cr \citep[e.g.,][]{Bergemann2010Cr} and can explain the decrease in the Cr/Fe ratio in the halo 
with decreasing metallicity. 
Cobalt has mixed contributions from SNe of type Ia and II, while  hypernovae are also viable channels \citep{Timmes1995,Kobayashi2006}. 
As we did for Cr, here we did  not correct  our Cr results for NLTE \citep{Bergemann2010Co}.
The values for [Co/Fe] in Pal~15 are entirely in line with those
in halo field stars, and  little variation in this element is seen in the other GCs.
\subsection{Neutron-capture elements (Sr, Ba, Eu)}
Abundances for three tracers of the neutron-capture processes were determined by using
spectral synthesis. 
Here the mean abundance and errors were estimated via the best matching synthetic fit for
each absorption feature used, weighted by the quality (in terms of S/N and continuum setting) of the respective lines.  
The resulting abundance ratios are  shown in Fig.~7, where we overplot the halo comparison samples from \citet{CJHansen2012} and \citet{Roederer2014}
and disk stars from \citet{Battistini2016}.
\begin{figure}[htb]
\centering
\includegraphics[width=1\hsize]{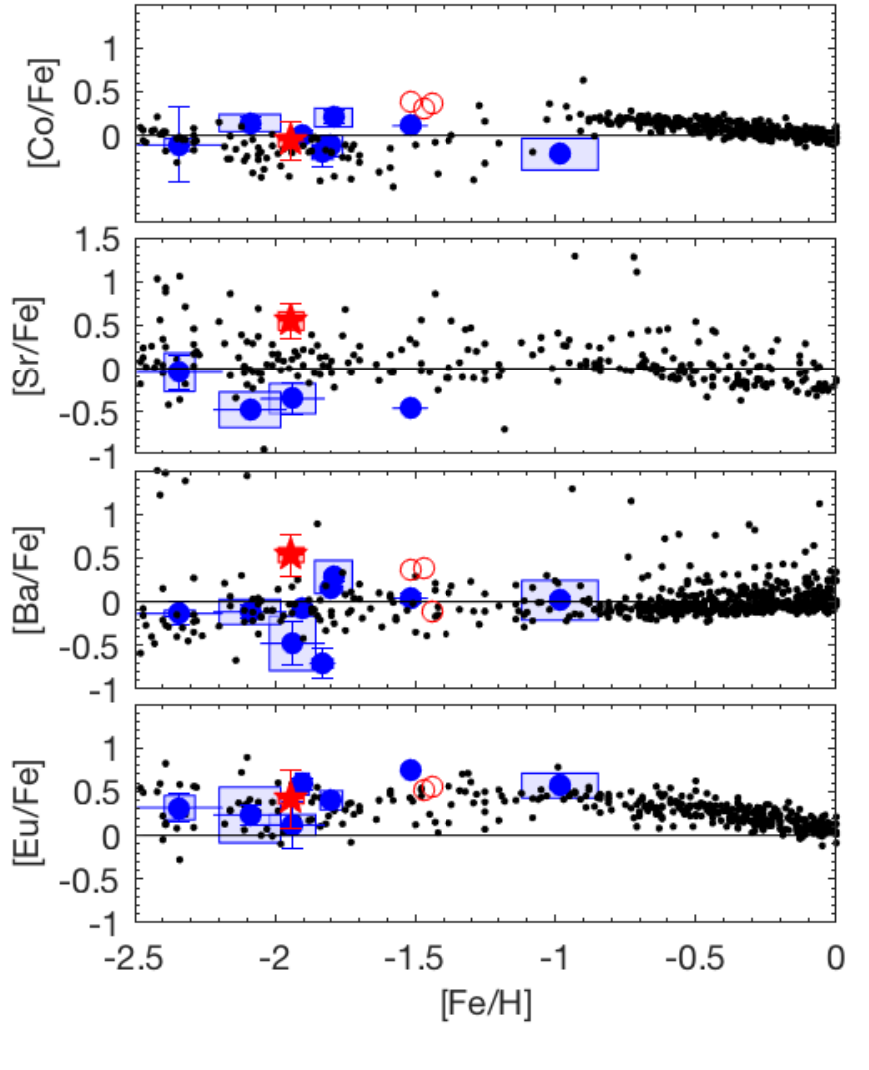} 
\caption{Same as Fig.~5, but for Co and the neutron-capture elements Sr, Ba, and Eu.}
\end{figure}

Strontium abundances were determined from the resonance line at 4215\AA, the usefulness of which even at low resolution has been demonstrated 
by, for instance, \citet{Stanford2010} and \citet{Koch2017ESO}. 
As above, we  did not correct our results for Sr for departures from LTE  as they are negligible at the stellar parameters of our stars \citep{CJHansen2013}.
As Fig.~7 indicates, the [Sr/Fe] ratio in Pal~15 grazes the higher values of the distribution of halo stars; however, the halo stars show a broad scatter at these 
low metallicities themselves. While the data in other GCs are sparse, the three outer halo GCs with measured Sr abundances are  characterized by 
solar to subsolar values. 

Our Ba measurements are based on synthesizing the typically well-described lines at 4554, 5853, 6141, and 6496\AA.
{Here we adopted the solar-scaled isotopic ratio, but we note that switching to pure $r$- or pure $s$-ratios 
\citep{Sneden2008} essentially leaves our results of these moderately metal-poor stars unchanged.}
The resulting mean abundance of $\sim$0.57 dex appears rather high in the Galactic context.
On the other hand, 
\citep{Andrievsky2009} estimated NLTE corrections on the order of 0.2 dex for stars with similar stellar parameters to those in our sample.
These values would correct our inferred Ba abundances downward to reasonable levels when compared to halo and GC stars. 

Finally, we  estimated Eu-abundances from the 6645\AA~line, which is very weak, however, leading to typical uncertainties at the 0.2--0.3 dex level. 
In turn, the blue 4129\AA~line yielded additional information and we report a visually weighted average of both features as our final Eu abundance in the following.
At $\sim$0.4 dex, the respective value is consistent with the majority of halo field stars and Galactic GCs with reported abundances. 
\section{Comparison with outer halo GCs beyond 20 kpc}
In the following, we  contrast our measurements in Pal~15 with the abundance ratios found in  other outer halo GCs;  we restrict the sample to the
objects at Galactocentric distances beyond $\sim$15 kpc. 
With few exceptions,
 all of these are moderately metal poor at an [Fe/H] of about $-1.5$ dex and most 
show $\alpha$-enhancements that are typical of the Galactic halo at these metallicities;  we discuss deviations in Sect.~5.1.
Combined with their, on average, younger ages \citep[e.g.,][]{Stetson1999,Marin-Franch2009} 
this argues in favor of them being accreted objects, adding to the complexity of the formation of the outer halo. 

Some of the reference GCs  in the literature claim to show broad spreads in some abundances past the key CN/CH or Na-O variations. 
In particular, proton-capture reactions in the progenitors of the second stellar populations tend to produce a mild Mg-Al correlation, while  
spreads in some $s$-process elements are also seen, pointing to specific channels of enrichment from asymptotic giant branch (AGB) stars \citep{Marino2015,Cordero201747Tuc}.
Furthermore, Fe spreads have been reported in about ten of the MW GCs studied in detail at high spectral resolutions.
To this end,  we show the abundance distribution of these clusters via boxplots, where the height and width of each box illustrates the observed
1$\sigma$ dispersion in a given abundance ratio and in the measured Fe abundance in the stellar samples (of typically a few to tens of stars). 
We note that the appearance as broad boxes does not imply that the GCs in question have significant intrinsic abundance spreads of 
that magnitude. Moreover, this illustrates the possible parameter space inhabited within the error bars. These error bars  in the figures
refer to the mean abundance error as stated in the literature. In this regard, error bars exceeding the box sizes indicate that the observed dispersions 
are mainly driven by observational errors, while those immured by the $\sigma$ box would suggest the presence of real intrinsic spreads. 
Individual cases are discussed below. 
\subsection{Individual outer halo GCs}
For three of the objects, abundance ratios are only available from co-added spectra. That is, individual, low S/N spectra of stars with known stellar parameters 
were combined into a master spectrum and were compared against a co-added synthetic spectrum with the same properties. This yields a mean abundance
ratio of the GC in question. However, as pointed out by \citet{Koch2017Pal5}, this method is highly insensitive to intrinsic abundance spreads in the GCs, 
and even the prominent light-element variations (e.g., in Na and O) are undetectable to the 0.6 dex level.
These very remote systems comprise 
Pal\,5 (R$_{\rm GC}$= 19 kpc; \citealt{Koch2017Pal5}),  
Pal\,4 (R$_{\rm GC}$= 111 kpc; \citealt{Koch2010}), and  
Pal\,14 (R$_{\rm GC}$= 77 kpc; \citealt{Caliskan2012}).
Pal\,3 (R$_{\rm GC}$= 96 kpc) has both co-added and individual measurements available, which yield consistent results \citep{Koch2009}. 

Among the objects with available high-resolution abundances from individual stars, 
the metal-rich ([Fe/H]=$-1$ dex) GC M\,75 (R$_{\rm GC}$= 15 kpc) 
shows an indication of a spread in its Fe-abundance typical of its luminosity.
Furthermore \citet{Kacharov2013} 
detected three generations of stars in abundance space, and  variations in the masses of AGB polluters, leading to a 
broad spread in $s$-process elements between stars of either population. 

NGC\,6426 (R$_{\rm GC}$$\sim$15 kpc) is one of the most metal-poor systems in the MW halo ([Fe/H]=$-$2.3 dex).
While only based on four stars, \citet{Hanke2017} suggested abundance spreads, though marginal, in several elements, 
adding detections of a correlation between Mg, Si, and Zn to the pool of the canonical (C,N,Na,O,Mg,Al) light-element variations. 

The metal-poor ([Fe/H]$\sim-2$ dex) NGC 5634 (R$_{\rm GC}$= 21 kpc) is a GC that has been associated with the disrupted Sagittarius (Sgr) dwarf galaxy based on position and dynamical arguments \citep{Law2010}, 
although its chemical abundances are indistinguishable from those of MW halo field stars \citep{Sbordone2015}. 
Here we used the extended chemical abundance measurements from \citet{Carretta2017}, who found no evidence of any significant abundance spread.
Similarly, Arp~2 (at 21 kpc; [Fe/H]=$-1.8$) is a purported member of the Sgr system. 
 \citet{Mottini2008} performed a high-resolution abundance study of this GC, although the small sample size of 
 two stars prevents a detailed investigation of the question of abundance spreads. 

NGC 5824 is an intriguing object at R$_{\rm GC}$= 26 kpc whose nature is still disputed. 
For instance, \citet{Mucciarelli2018} measured a systematic difference in iron abundance between their 
AGB and RGB samples  at the 0.1 dex level, which is, however, easily  explained as  measurement uncertainties. 
The same study ascertained the presence of an ``extreme'' Mg-Al anti-correlation as typically seen in metal-poor and/or very massive clusters, and driven by the self-enrichment via
massive AGB stars. 
The authors thus ruled out the possibility that NGC~5824 is the remnant of a disrupted dwarf galaxy.  This is in stark contrast to 
previous morphological evidence \citep[e.g.,][]{Kuzma2018} and the recent finding of  \citet{Yuan2019} who, kinematically, identified this GC with the nuclear star cluster of the Cetus stream. 
In our figures, we include  the comprehensive abundance measurements from \citet{Roederer2016NGC5824}, who reported on further abundance spreads in this cluster, 
but we note the overall more metal-rich mean found by those authors. 

NGC 5694 (R$_{\rm GC}$= 29 kpc) is of similar interest: for this GC,  \citet{Mucciarelli2013} find 
remarkably low abundances of the $\alpha$-elements and several neutron-capture elements.
While Na and O show the expected broad ranges due to the evolutionary anti-correlation, 
no other abundance spreads have been found. Based on the overall  low mean abundances, these authors 
confirmed the hypothesis of \citet{Lee2006} that NGC~5694 is of extragalactic origin, which is also supported by the presence of an extratidal halo around this GC \citep{Correnti2011}.

Next, four stars in NGC 7492 (R$_{\rm GC}$= 25 kpc) have been analysed by \citet{Cohen2005}, who
find this outer halo GC to be chemically indistinguishable (including the neutron capture elements)  from its inner halo counterparts such as M3 or M13.  
With the additions of the even more remote Pal 3 and Pal 4, which share the same chemical similarities  \citep{Koch2009,Koch2010}, this 
bolsters the view of a homogeneous chemical history of the inner and outer halos at least through the  formation epochs of these old systems, 
as witnessed by these metal-poor GCs. 
Furthermore, \citet{Cohen2005} find no abundance spreads in this object safe from the expected variations in Na and O.  
 
NGC~6229 (R$_{\rm GC}$=30 kpc) is a rather metal-rich system ([Fe/H]$\sim$-1.1) with a 
modest dispersion of 0.06 dex in iron \citep{Johnson2017}. Moreover, these spreads also permeate into the heavier elements. 
For instance, \citet{Johnson2017} find that
2 of 11 stars in their sample show enhanced La and Nd abundances that possibly correlate wth Fe. 
Furthermore, the stars in this GC show, overall,  systematically lower Na  and Al abundances, suggesting an accretion origin, possibly as the core of a disrupted dwarf galaxy. 
Given the similarity of  NGC~6229  to two other GCs (M75 and NGC~1851) with regard to many abundance ratios and other characteristics, 
\citet{Johnson2017} dubbed this system and its brethren an Fe-complex cluster, whose members  are characterised by prolonged enrichment histories.

The most remote GC with reported individual high-resolution abundances is NGC 2419  (R$_{\rm GC}$= 86 kpc). It has been characterized as a  bizarre object that not only 
exhibits a remarkable metallicity spread, but also significant spreads in certain light elements that have, to date, not been seen in other GCs \citep[e.g.,][]{Cohen2012}.
In particular, the distribution of Mg abundances is 
bimodal and strong variations in K and Sc have been found. 
Iron and calcium, on the other hand, are well behaved and show no significant spreads whatsoever.
This cluster remains puzzling: while \citet{Cohen2012} point out the resemblance of the Mg depletions with those seen in dwarf spheroidal galaxies
and the lack of any nucleosynthetic sites that could simultaneously explain all of the abundance variations seen in this object, 
the lack of a significant Fe spreads  contradicts this unique explanation.  
NGC~2419 remains unique in that it is like no other GC, and it may (or may not) be the core of an accreted dwarf galaxy. 
In the following figures, we use the  abundance data from \citet{Mucciarelli2012} and \citet{Cohen2012} for the heavier elements.
\subsection{Comparison with Pal~15}
Several GCs share individual abundance ratios with Pal~15, although none has the exact same pattern. 
In Fig.~8 we thus compare the abundance pattern of Pal~15 with those in the outer halo GCs discussed above. 
Here we note that, strictly speaking, our abundance ratios for certain elements should be corrected   for the effects of NLTE.
However, our targets are  mainly red giants with small corrections. More importantly, here we are 
 comparing our results with GCs of similar metallicity and targets of similar stellar parameters. Thus,
 all objects experience the same order of corrections, thereby minimizing systematic differences
 in the present comparative analysis.
\begin{figure*}[htb]
\centering
\includegraphics[width=0.49\hsize]{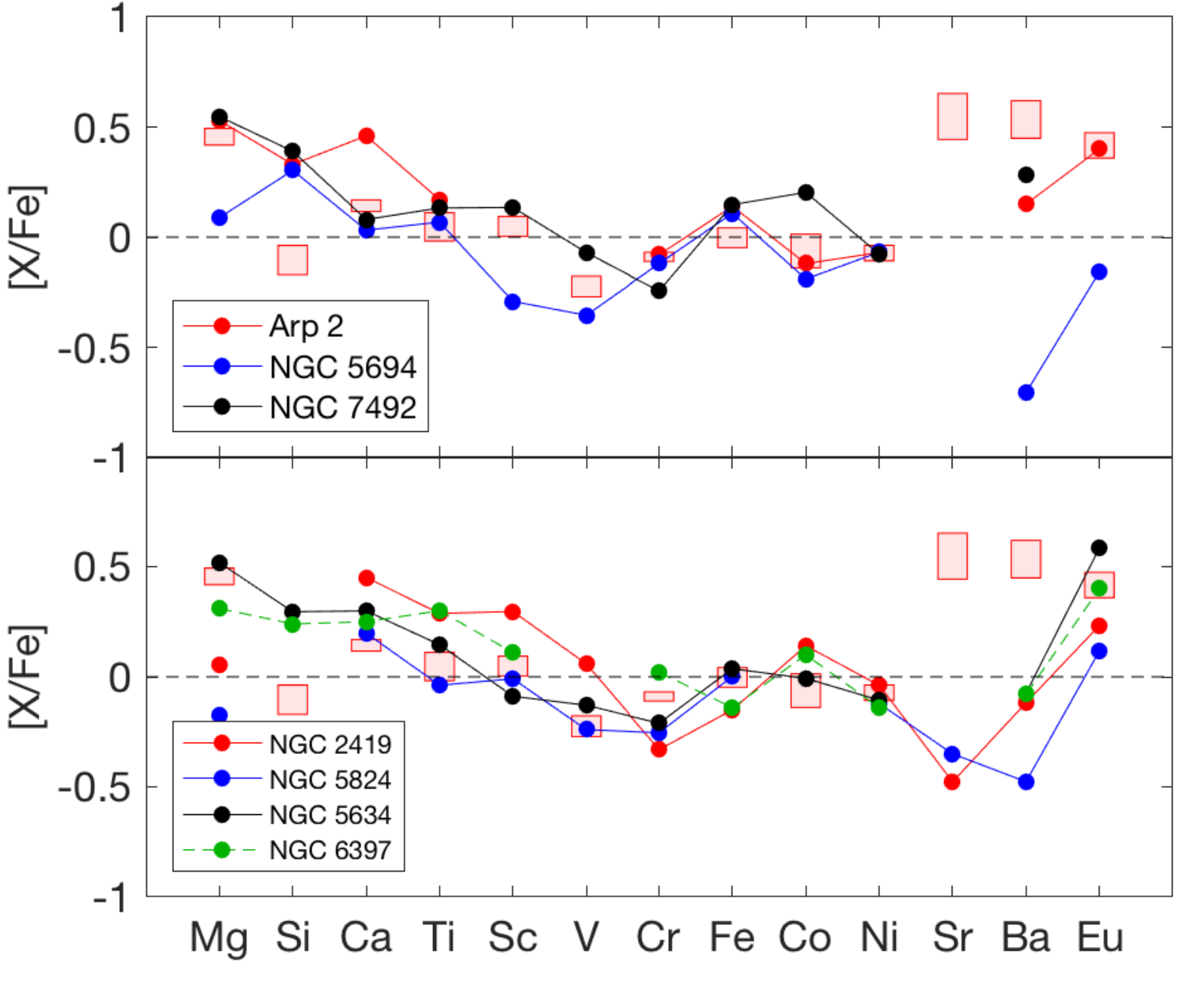} 
\includegraphics[width=0.49\hsize]{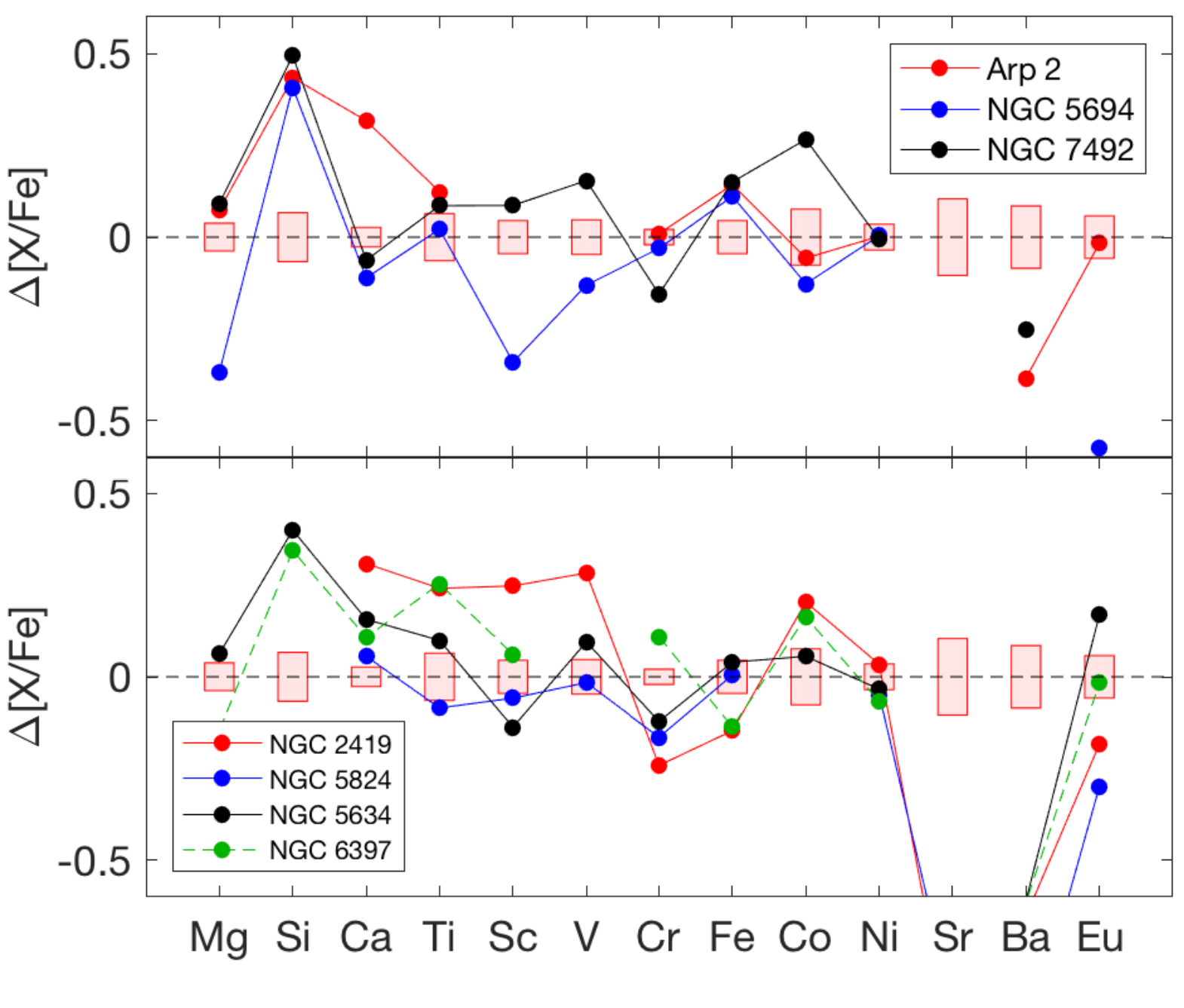} 
\caption{Comparison of Pal~15 (red boxes, scaled to the respective abundance scatter) with seven GCs at similar metallicities. 
{The left panels are for abundance ratios, whereas the right ones show the difference [X/Fe]$_{\rm GC} - $[X/Fe]$_{\rm Pal 15}$.}
To help readability, the figures are split into two panels each;  the top panels show the three objects at $-$1.8 dex, and 
the bottom panels show the remaining more metal-poor reference GCs. For iron, we show [Fe/H] relative to the Pal~15 mean.}
\end{figure*}

Among the reference objects considered here, NGC~7492 shows a good resemblance, {in a statistical sense},  to the elements studied here. 
In particular, for 5 out of the 11 elements in common, the patterns of these two GCs agree to better than 0.1 dex. 
Among the elements that deviate more are those elements mainly derived from weak lines (Si, V); Cr, for which NLTE corrections may 
be relevant; and Ba.
{As  \citet{Cohen2005} characterized  NGC~7492 as a typical halo cluster, the resemblance mentioned above 
suggests that  Pal~15 is equally ordinary.}
While it appears tempting to associate NGC~7492 and Pal~15 as an Fe-complex cluster because they  
share the same chemical properties \citep{Johnson2017}, we note that neither of these objects shows any significant spread in 
the elements considered here. Their similarity  attests to a certain degree of homogeneity in the outer halo population, at least as sampled by these GCs. 

{As a final comparison object we tested our results against the cluster 
NGC~6397, which  has a  similar metallicity \citep{Koch2011}, but is located in the inner halo at a mere R$_{\rm GC}$=6 kpc. 
Here, we make a comparison with the NLTE abundances from \citet{Lind2011NGC6397}. As the green dashed line in Fig.~8 shows, 
 one-third of the elements in common between the two outer and inner halo GCs agree to better than 0.1 dex, emphasizing the 
complexity and variety of GCs that are not necessarily dependent on 
the inner or outer halo location \citep{Milone2017}. 
}%
\section{Discussion}
Our abundance analysis of three red giants in the outer halo GC Pal~15 has {revealed no strong evidence of it being an accreted object, 
but rather suggests that it is an ordinary metal-poor halo cluster.} 
Its similarity to another outer halo GC, NGC~7492, that has been shown to closely follow the 
outer halo population, indicates the chemical homogeneity of this Galactic component with respect to its non-accreted population.  
Taking Mg as a representative $\alpha$-element, we do not see a depletion that would be 
a key signature of an extragalactic accretion origin. {
Lower values are mainly seen in Ca and Ti, which are, however, still compatible with the 
lower boundary of halo field and other GC stars. In order to resolve the seeming ambiguity between the different $\alpha$-elements, 
spectra at higher resolution of a larger sample of stars are needed, which is  challenging given the GC's faintness.}

The Fe-peak elements are slightly subsolar to solar and well-behaved, which also holds for the few neutron-capture species we sampled. 
This is bolstered by the absence of any significant spread in any of the abundances studied. 
Pal~15 is a fairly faint system ($M_V=-5.5$). Other objects of similar low luminosities and comparably old ages 
are known to have produced a Na-O anti-correlation \citep{Bragaglia2017,Bastian2018}. {While we see tentative evidence 
of a second generation of Na-enhanced stars in Pal~15,  this
needs to be consolidated from higher resolution studies of a larger sample of stars.}

The remainder on the list of known outer halo GCs beyond 20 kpc \citep{Harris1996} without 
any published  abundances from high-resolution spectroscopy, to the best of our knowledge, is surprisingly long:  
NGC 4147 (21 kpc), 
Pal~13 (27 kpc), 
AM~4 (28 kpc), 
Whiting~1 (35 kpc), 
Pal~2 (35 kpc), 
NGC 7006 (39 kpc),  
Pyxis (41 kpc), 
Koposov~1 and 2 (49 and 42 kpc), 
Eridanus (95 kpc), and 
AM~1 (125 kpc).
In some cases, low-resolution metallicity estimates are available from the near-infrared calcium triplet \citep[e.g.,][]{Suntzeff1985}, and 
for others the presence of multiple populations and light-element variations has been ascertained from photometric and low-resolution data 
\citep[e.g.,][]{Harbeck2003NGC7006,Gerashchenko2018}. 
Obtaining the full chemical abundance information for these systems is imperative for 
disentangling the formation history of outer halo and the occurrence of abundance variations as a function of GC mass and environment.  
For the most remote objects, this is hampered by the sparseness of the GCs and faintness of the accessible red giants, with the 
brightest member stars being fainter than V=18 mag \citep{Hilker2006}, which renders detailed abundance measurements time-intensive.
\end{CJK}
\begin{acknowledgements}
A.K. acknowledges financial support from the Sonderforschungsbereich SFB 881 
``The Milky Way System'' (subprojects A03, A05, A08) of the DFG. 
The authors are grateful to C.J. Hansen and M. Hanke for the helpful discussions
and to the anonymous referee for the constructive suggestions. 
This work has made use of the SP\_Ace spectral analysis tool version 1.2.
The data presented herein were obtained at the W.M. Keck Observatory, which is operated as a scientific partnership among the 
California Institute of Technology, the University of California, and the National Aeronautics and Space Administration. The Observatory 
was made possible by the generous financial support of the W.M. Keck Foundation.
The authors wish to recognize and acknowledge the very significant cultural role and reverence that the summit of Mauna Kea has 
always had within the 
indigenous Hawaiian community.  We are most fortunate to have the opportunity to conduct observations from this mountain. 
 \end{acknowledgements}
\bibliographystyle{aa} 
\bibliography{ms} 

\end{document}